\documentclass[a4paper,prl,twocolumn,showpacs]{revtex4-1}
\usepackage[dvips]{graphicx}
\usepackage{amsmath,amssymb}
\usepackage{dcolumn}
\usepackage{bm}
\usepackage{color}

\begin{document}
\bibliographystyle{apsrev}

\title{Inverse simulated annealing for the determination of amorphous structures}

\author{Jan H. Los} 
\author{Thomas D. K\"uhne}
\email{kuehne@uni-mainz.de}
\affiliation{Institute of Physical Chemistry and Centre for Computational Science, 
Johannes Gutenberg University Mainz, Staudinger Weg 7, D-55128 Mainz, Germany}

\date{\today}

\begin{abstract}
We present a new and efficient optimization method to determine the structure of 
disordered systems in agreement with available experimental data. 
Our approach permits the application of accurate electronic structure calculations 
within the structure optimization. The new technique 
is demonstrated within density functional theory by the calculation of a 
model of amorphous carbon. 
\end{abstract}


\maketitle
Amorphous solids 
can be produced from almost any chemical system and 
are of great interest due to their large variety of technologically important 
applications. In addition to conventional silicate glasses, they are, for example used 
in optical waveguides (oxides), plastics (organic polymers), solar cells 
(semiconductors), biomaterials (amorphous metals), xerography 
and non-volatile memory devices (chalcogenides), to name but a few 
\cite{Zallen,Elliott}. {Nevertheless, finding their atomic scale structure is 
still a major challenge in material science \cite{Zachariasen,Elliott,Jansen} 
due to the absence of lattice periodicity and long-range order characteristics
of a crystalline solid. Many sophisticated modeling techniques from the field of crystal structure prediction are based on 
searching the global minimum in the energy landscape for periodic structures 
\cite{Maddox,Deaven,WalesScheraga,Martonak,MinHopping,USPEX,Trimarchi,
Catlow,Wehmeyer}. 
However, an amorphous solid does not correspond to a global, but to a local energy minimum, which is energetically low enough to stabilize the structure against alternative packings and exhibits desirable target properties.


The most commonly applied computational technique to obtain the amorphous 
structure is to slowly quench it from the melt by Monte Carlo (MC)- or 
Molecular Dynamics (MD)-based Simulated Annealing (SA) \cite{SA}. 
However, the lack of exploitable symmetry and, therefore, large number 
of degrees of freedom, 
require the cooling to be conducted as slowly as possible to determine an 
approximation of the amorphous structure, and is, therefore, computationally 
very demanding. This is even more pronounced in conjunction with accurate 
\textit{ab-initio} electronic structure techniques, in spite of 
significant progress in recent years \cite{CarParrinello, KuhneParrinello}, 
allowing for satisfactory structure determinations 
\cite{Pasquarello,Marks,CaravatiAPL,*CaravatiPRL,*CaravatiJPCM1, *CaravatiJPCM2, Camellone}.


Instead of performing an elaborate calculation to obtain an approximate 
amorphous model and to assess \textit{a posteriori} how well it matches 
the experiment, McGreevy and coworkers demonstrated that it can be beneficial 
to reverse this procedure, hence the name Reverse Monte Carlo (RMC) 
\cite{McGreevy1,McGreevy2}.
Contrary to energy-based minimization techniques this method aims at 
directly modeling the structure without invoking any computationally expensive 
potential energy calculation, using only available experimental data. 
Specifically, the available experimental data are reproduced simply 
by minimizing a function of the form 
\begin{eqnarray}
\label{F_RMC}
  \mathcal{F(\mathbf{R})} = 
  \sum_p w_p \left( \chi_{p}(\mathbf{R}) - \chi^{exp}_{p} \right)^2
\end{eqnarray}
under variation of the atomic positions $\mathbf{R}= \{ {\mathbf r_i} \}$ 
using the Metropolis Monte Carlo method \cite{Metropolis}. 
In Eq.~\ref{F_RMC}, $\chi_{p} (\mathbf{R})$ and $ \chi^{exp}_{p}$ are the 
calculated and experimental values, respectively, of a property $p$, while $w_p=1/\sigma_p^2$ 
is a weight factor and $\sigma_p$ is the experimental uncertainty for the 
corresponding property. 

Even though $p$ can, in principle, be any arbitrary property, in practice, 
only geometric quantities, obtainable from Neutron or X-ray scattering 
data such as the structure factor or the pair correlation function for which 
$ \chi_{p}(\mathbf{R}) $ can be evaluated easily and fast, are employed.
In particular, typically no electronic quantities based on accurate 
electronic structure calculations are utilized, which would otherwise be 
computationally unfeasible. While on the one hand, RMC allows for an 
efficient and routine modeling of rather complex disordered structures, on the other hand, 
the resulting models are not necessarily physically sensible. 
It is, therefore, good practice to circumvent that as much as possible by 
imposing specifically selected constraints \cite{McGreevy2, Kugler}. Although, eventually, this often 
leads to rather pleasing results, this may not be the case when studying unknown 
systems where good constraints are not known from the outset. 
In addition, since the atomic configuration in RMC is not relaxed into a 
local energy minimum, the resulting structure is not necessarily stable.

The inverse design technique of Franceschetti and Zunger allows one to, 
at least partially, circumvent the shortcomings just mentioned by determining 
the crystal structure based on electronic structure properties, which are 
rather sensitive with respect to the atomic positions. In their method, an inner local geometry optimization is performed in each optimization step to relax the structure \cite{Franceschetti}. 
However, for the sake of efficiency, the latter is conducted using an empirical valence force field only \cite{Keating}. 
Furthermore, in order to facilitate the 
calculation, they confined themselves to highly symmetric structures on a given 
crystal lattice.

In this work, we improve upon the existing approaches by 
proposing a novel and efficient method, which we call Inverse Simulated Annealing (ISA). 
This method combines the global minimization of a linear combination consisting of various 
geometric and electronic properties with structure relaxation to determine 
an amorphous solid in best agreement with available experimental data. 
Specifically, this is achieved by adding the potential energy $U(\mathbf{R})$ to the objective function of Eq.~\ref{F_RMC}, 
and employing a modified hybrid Monte Carlo (HMC)-based SA scheme to minimize it. 
We will demonstrate that the present method is efficient enough to be
applicable in conjunction with accurate electronic structure calculations, 
and in this way allows to routinely determine the amorphous structure.

In the following, we will confine ourselves to effective single-particle theories, 
such as density functional theory (DFT) \cite{DFT}. Hence, the modified objective 
function to be minimized reads as: 
\begin{eqnarray}
\label{Utilde}
\tilde{U}(\mathbf{R}) &=& U(\mathbf{R}) 
+ \sum_p w_p \left( \chi_{p}(\mathbf{R}) - \chi^{exp}_{p} \right)^2 \nonumber \\
&+& \sum_q w_q \left( \xi_{q}\left[ \mathbf{R}, \{\psi_{i}\} \right] - \xi^{exp}_{q} 
\right)^2,
\end{eqnarray}

where $\tilde{U}(\mathbf{R})$ is a fictitious and $U(\mathbf{R})$ the potential energy, 
as obtained by DFT, while 
$\xi_{q}\left[ \mathbf{R}, \{\psi_{i}\} \right]$ and $\xi^{exp}_{q}$ 
are the computed and experimental values, respectively, of an electronic quantity $q$. 

In minimizing $\tilde{U}(\mathbf{R})$, we take advantage of 
the fact that by using Eq.~\ref{Utilde}, the accessible phase space 
is substantially reduced and restricted to energetically low-lying 
configurations. 
In other words, even though the dimensionality of the phase 
space is equally vast, the optimization is guided in a funnel-like fashion 
towards the minimum of $\tilde{U}(\mathbf{R})$. Obviously, in spite of that, 
we still need a global optimization method to minimize Eq.~\ref{Utilde} that 
is efficient enough to enable the calculation of $U(\mathbf{R})$ at the DFT level of theory. 
The fact that the derivatives 
of some of the properties in Eq.~\ref{Utilde} with respect to $\mathbf{R}$ are 
not directly available and may not even exist due to possible discontinuities, 
immediately suggests a MC-based minimization procedure \cite{Duane}. 
The development of such a technique is therefore an essential part of the present work. 

For the purpose to minimize Eq.~\ref{Utilde}, while at the same time using as few as possible 
electronic structure calculations, we propose here a novel 'fuzzy' HMC-based SA scheme within 
the NVE instead of the more common NVT ensemble that consists of only a single 
modified MD step. In comparison to standard MC- or MD-based SA techniques, 
we found that this technique performs particularly well as a minimization method, 
as will be shown.
The positions and velocities of all atoms in each trial move are varied according 
to a slightly modified velocity-Verlet algorithm:
\begin{eqnarray}
\label{verlet}
\left\{ \begin{array}{l}
\displaystyle
\vspace*{0.2cm}
{\bf r}'_i = {\bf r}_i + {\bf v}_i dt + \frac{1}{2} 
\frac{\tilde{\bf f}_i}{m_i} dt^2 \\
\vspace*{0.2cm}
{\bf v}'_i = C \left( {\bf v}_i + \frac{1}{2} 
\left( \frac{\tilde{\bf f}_i}{m_i} + \frac{\tilde{\bf f}'_i}{m_i} \right) dt \right)
\displaystyle
\end{array} \right.,
\end{eqnarray}

where ${\bf v}_{i}$ are the ionic velocities, $m_{i}$ the nuclear masses, 
$dt$ a randomly chosen time step from an uniform distribution within the interval [0, $dt_{max}$], 
while the prime superscripts are used to indicate quantities of the new 
(trial) configuration. The forces $ \tilde{\bf f}_i $ are the best possible 
estimate for 
$- \partial{\tilde{U}}/\partial{\tilde{\mathbf{r}}_i}$, 
i.e. omitting the contributions from those terms in the sums of Eq.~\ref{Utilde} 
for which no derivatives are directly available. This and the presence of 
a maximum time step $dt_{max}$, which is in general much larger than in standard 
MD and continuously adjusted to obtain an acceptance rate of about 50~\%, 
is why we call our modified HMC algorithm 'fuzzy'. 
In order to ensure that the total energy is conserved, in Eq.~\ref{verlet} we have introduced 
an additional prefactor denoted as $C$, which chosen in such a way that 
$1/2 \sum_i^N m_i | {\bf v}'_i |^2 = K' = E - U'$ holds, where $K'$ is 
the kinetic energy of the system of the proposed trial configuration. 
Within the NVE ensemble, the probability of acceptance of a trial 
move is given by \cite{Ray}
\begin{eqnarray}
\label{probNVE}
P = \min{ \left( 1, \left( \frac{ E - U' }{ E-U } \right)^{3N/2 -1} \right) },
\end{eqnarray}
where $N$ is the number of atoms.

As already mentioned, the present approach differs from the standard HMC 
algorithm in the fact that the NVE instead of the usual NVT ensemble is employed. 
Furthermore, only a single MD step is taken in each HMC step and the velocities 
are not randomly re-initialized thereafter. The necessary random element in our HMC method comes from the randomly chosen, variable time step $dt$ instead. 
Whenever a HMC move is accepted, the positions and velocities are updated as 
$( {\bf r}_i , {\bf v}_i ) = ( {\bf r}'_i , {\bf v}'_i )$, just as in normal MD. 
Otherwise, if an HMC move is rejected, then one possibility is to maintain 
$ ( {\bf r}_i , {\bf v}_i )$, in which case no update is required. We will 
denote this straightforward version of our method as fHMC-NVE. However, 
regarding the efficiency of the minimization procedure, it is desirable to 
design an algorithm that combines a large time step with a high acceptance rate. 
It appears that an improvement in this direction is obtained by maintaining 
the velocities of the rejected configurations, i.e. by updating according 
to $ ( {\bf r}_i , {\bf v}_i ) = ( {\bf r}_i , {\bf v}'_i ) $ after a rejection. 
In this modified algorithm, indicated hereafter as mfHMC-NVE, the velocities 
are gradually turned in the direction of the forces upon repeated rejections. 
As a consequence, the acceptance probability for large displacements 
(i.e. large $dt$) increases, since the displacements become more and more 
parallel to the forces, i.e. the direction of decreasing potential energy. 

\begin{figure}[htb]
\vspace*{0.0cm}
\includegraphics[width=8cm,clip]{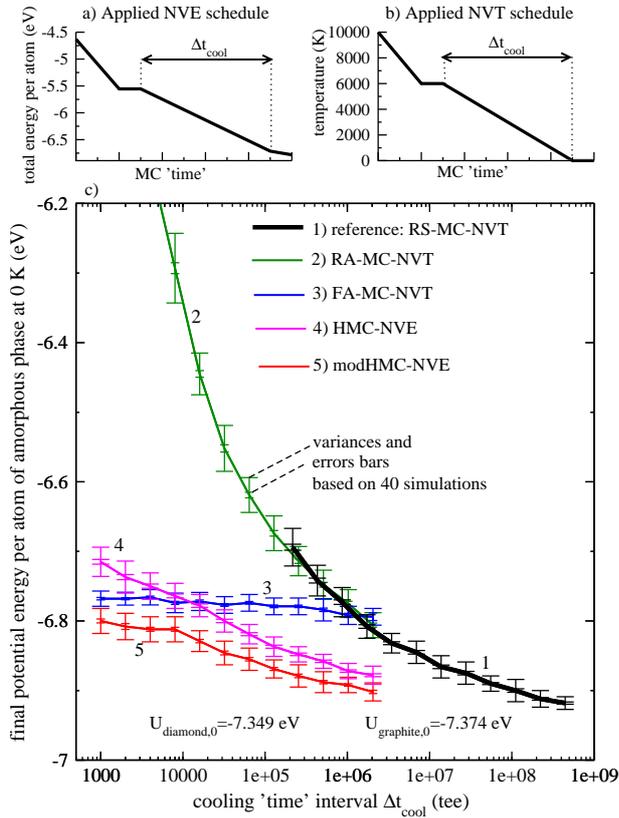}
\caption{\label{fig:E_tq} Comparison of the average over the final potential 
energies at 0~K of amorphous carbon as generated by the various minimization 
method as a function of the quenching time $ \Delta t_{cool} $. The averages 
are based on 40 independent simulations, 
allowing for the calculation of variances and error bars as indicated.
Note the logarithmic scale for the 'time' axis.}
\end{figure}

To assess the performance of our HMC-based minimization technique, 
we have applied it to carbon using the empirical LCBOPII potential \cite{Los}. 
This bond order potential has been shown to accurately describe many carbon 
phases including the disordered, liquid phase within a whole range of different 
densities \cite{Ghiringhelli}. We have selected a system consisting of 216 atoms 
within in a cubic simulation box with periodic boundary conditions, which 
corresponds to a density of $ \rho = 3.1 $ g/cm$^3$, which is in close agreement 
with the experimentally determined density of amorphous carbon \cite{Gilkes}. 
For the sake of simplicity, in these simulations, meant to test and compare the performance of different 
minimization techniques, momentarily only the potential energy is minimized. 

The applied total energy schedule as a function of the (fictitious) MC 'time', 
is schematically shown in Fig. \ref{fig:E_tq}a. Starting at a high total 
energy $E= -1000$~eV, to create a well disordered liquid phase, 
the schedule includes a liquid equilibration
period at constant $ E= -1200$~eV, after which the system is cooled down 
linearly to $ E = -1450$~eV during a 'time' interval $\Delta t_{cool}$. 
After that, the system is relaxed in a relatively short quench by further 
decreasing $ E $ to a value close to the final potential energy 
$ U_{f,0}(\mathbf{R})$. Note that the instantaneous temperature of the 
system can be deduced from $K = (3/2) N k_B T = E - U$, which implies 
$T = 2 ( E - U )/ ( 3 N k_B )$, so that $ T \rightarrow 0 $~K for 
$ E \rightarrow U_{f,0} $. 

The results for the average, final potential energy per atom at 0~K, 
$ U_{f,0}(\mathbf{R})/N $, as a function of the cooling 'time' interval 
$\Delta t_{cool}$ in units of total energy evaluations (tee), based on 
40 independent simulations, are shown in Fig.~\ref{fig:E_tq}c and compared 
to the results from other, more standard minimization techniques. 
These include the reference, a random single atom displacement MC method 
within the NVT ensemble, indicated as RS-MC-NVT, and two all atom MC methods 
within the NVT ensemble: the RA-MC-NVT method with completely random, 
simultaneous displacements of all atoms and the FA-MC-NVT method, where 
the displacement of each atom is a mixture of a random vector and the force 
on that atom with a mixing coefficient chosen such that the efficiency 
is maximized. The applied temperature versus the MC 'time' for 
these NVT simulations is schematically given in Fig.~\ref{fig:E_tq}b.

As can be seen in Fig.~\ref{fig:E_tq}c, the behavior of the RA-MC-NVT technique and the reference is essentially 
identically, which suggests that in the present case it is insignificant if either all or a single atom is randomly displaced. 
Nevertheless, the straightforward inclusion of nuclear forces in the FA-MC-NVT approach leads to an optimization 
scheme that can easily get trapped in a local minimum and is hence not competitive. On the contrary, in the (m)fHMC-NVE 
method this is circumvented by the interplay of $dt$ and $C$. 
On average $dt_{max}$ is relatively large, i.e. typically about one order of magnitude larger than in a conventional 
MD simulation for carbon and remains approximately  constant during the annealing. In this way, the available 
gradient information is rather well exploited. However, upon rejections the decrease of $dt$ is counterbalanced by 
$C$ to conserve the instantaneous total energy and therefore prevents the system to be trapped in a local minimum. 
In the end, employing the mfHMC-NVE method, the same potential energy than using the RS-MC-NVT approach is realized, 
though with a two orders of magnitude shorter cooling time. Comparing the final potential energies with the ground state 
energies of diamond (-7.349 eV/atom) and graphite (-7.374 eV/atom), it is apparent that the eventual structures correspond 
to amorphous carbon, whose energies are about 0.4 eV/atom above the corresponding ground state.

As already mentioned, the mfHMC-NVE method shows the best performance regarding its 
ability to find low energy states. On the other side, it is feasible to do much longer simulations (in terms of tee) 
with the RS-MC-NVT method than using the 
other techniques, because the re-evaluation of the total energy 
after the displacement of one single atom is relatively fast for the empirical 
LCBOPII potential; this is due to the intrinsic local dependencies of the energy 
contributions in such potentials. Since the curve for RA-MC-NVT lies on top 
of that of the RS-MC-NVT technique, the latter is to be preferred whenever 
updating the total energy for single atom move is faster than for an all 
atom move. 

\begin{figure}[htb]
\vspace*{0.0cm}
\includegraphics[width=9cm,clip]{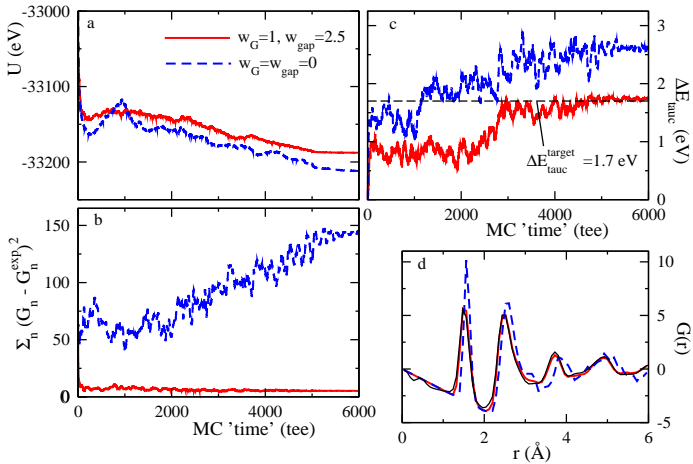}
\caption{\label{fig:fsimcomp} Evolution of the (a) potential energy, 
(b) the sum of the squared residuals of the RDF and (c) the tauc-gap 
as a function of time during the optimisation using the mfHMC-NVE technique 
to determine the structure of amorphous carbon. The solid line denotes 
our novel simulation method, while the conventional SA approach is depicted 
by the dashed line. The comparison of the corresponding $G(r)$, as obtained 
using both techniques, with the experimental one \cite{Gilkes}, given by the solid line, 
is shown in Fig.~\ref{fig:fsimcomp}d. Due to the nearly perfect agreement, the experimental 
curve is almost completely covered by the results from the present method (red line).}
\end{figure}

To illustrate our novel method, we apply the 
mfHMC-NVE method to minimize Eq.~\ref{Utilde} for amorphous carbon
at the density functional level of theory (DFT). Therein, $U(\mathbf{R})$ 
is the total energy from DFT supplemented by the reduced 
radial distribution function $G(r)$, derived from scattering data, and the 
optical Tauc gap $ \Delta E_{tauc}$ for amorphous phases \cite{Tauc}. 
Hence, in this case, Eq.~\ref{Utilde} takes the form:
\begin{eqnarray}
\label{Utilde_aC}
\tilde{U}(\mathbf{R}) &=& U(\mathbf{R}) +  w_G 
\sum_n ( G_n(\mathbf{R}) - G^{exp}_n )^2 \nonumber \\
&+& w_{gap} (\Delta E_{Tauc}(\mathbf{R}) - 
\Delta E_{Tauc}^{target} )^2,
\end{eqnarray}
where $G_n(\mathbf{R}) = G ( r_n )$ denotes a discretized representation 
of $G(r)$, which is defined as $G (r) = 4 \pi r ( c(r) - c_0 )$, 
with $c(r)$ the average (number) density of atoms at a distance $r$ 
and $c_0$ the overall density. To obtain a smoothened $G(r)$, allowing 
for the calculation of analytical force contributions that were included in the present simulations, 
we have computed it for any $r = r_n$ on a grid with a spacing of 0.01~\AA ~between the grid points as:
\begin{eqnarray}
\label{Gr}
G(r) = \frac{1}{ r \Delta r } \frac{1}{N} \sum_{i,j} 
\int_{r-\Delta r/2}^{r+\Delta r/2} P_{ij} (r') dr' - 4 \pi r c_0
\end{eqnarray}
where $P_{ij} (r)$ is a Gaussian-shaped
polynomial of degree 4 within the open interval 
$(r_{ij} - \Delta r , r_{ij} + \Delta r)$ and $P_{ij} (r ) = 0$ 
otherwise, with $P_{ij}$ and $dP_{ij}/dr$ being continuous at 
$r = r_{ij} \pm \Delta r$, $ \int P_{ij} (r) dr = 1$ 
and $r_{ij}$ the interatomic distance between atom $i$ and $j$. 
The values reported for the experimental gap of amorphous carbon 
vary between 1.0~eV and 2.5~eV, possibly depending on the particular 
sample \cite{Marks}. Therefore, we have taken an 
intermediate target value equal to $\Delta E^{target}_{tauc} = 1.7$~eV for our simulation. 
However, in the present study, we have neglected the gradient of the Tauc gap term with 
respect to $\mathbf{R}$ in the analytic expression of the forces. Nevertheless, using finite 
differences, it is straightforward to include them, although at the price that the computation 
becomes at least a factor of $3N$ times more expensive. Further details on the on-the-fly 
calculation of the Tauc gap are discussed in the Appendix.

Even though the values of the weight factors $w_G$ and $w_{gap}$ have some 
importance, their impact is relatively small. In principle they should 
be chosen as small as possible and just large enough to get a good agreement 
with the experimental data. In the present simulation we have used  
$w_G = 1$ and $w_{gap} = 2.5$. In general, the value $w_G$ should be chosen in such 
a way that, in the beginning of the simulation at high temperature, $\sum_n ( G_n(\mathbf{R}) - G^{exp}_n )^2$ 
is on the same order of magnitude than the thermal energy $\frac{3}{2}Nk_{B}T$. 
In contrast, the parameter $w_{gap}$ can be selected to be considerable smaller 
than $3 (\Delta E_{Tauc}(\mathbf{R}) - \Delta E_{Tauc}^{target} )^2 / 2Nk_{B}T$. 


We have linked our code 
to the CP2K suite of programs to compute the necessary total energies and forces \cite{VandeVondele}. The DFT calculations 
were performed using the Perdew-Burke-Ernzerhof (PBE) exchange correlation 
functional \cite{PBE} and norm-conserving Goedecker-type pseudopotentials 
\cite{Goedecker}.
The total energy schedule applied included an equilibration at 
$ E= -32950$~eV for 1000~tee, followed by a cooling from $ E=-32950$~eV 
to $ E=-33175$~eV during 4000~tee and a final run of length 1000~tee 
during which $E$ is further lowered to get as close as possible to 
$ \tilde{U}_{f,0} $. 

The results of such simulations using our novel method, with and without 
the experimental constraints, are presented in Fig.~\ref{fig:fsimcomp} 
and are compared to a conventional HMC-based SA simulation. 
The improved overall agreement with the underlying experimental data is apparent, as shown in Figs.~\ref{fig:fsimcomp}c and \ref{fig:fsimcomp}d.

We conclude by noting that our novel method in conjunction
with an appropriate minimization procedure has a wide domain of applicability, 
not limited to amorphous phases.
We wish to specifically highlight that the present scheme can be 
directly applied to any other disordered system, such as liquid water \cite{Kuehne2009, *Kuehne2011, *PascalWater, *Kuehne2013} or, including the NMR chemical 
shift \cite{MauriNMR1, *MauriNMR2, SebastianiParrinelloNMR, 
*SebastianiParrinelloWaterNMR}, 
to determine the structure of proteins and nucleic acids.
Further improvement of the method and the minimizer 
will be presented elsewhere.

\begin{acknowledgements}
We would like to thank A. Zunger for fruitful discussions, 
D. Richters for critical reading the manuscript as well as the IDEE project of 
the Carl-Zeiss Foundation and the Graduate School of 
Excellence MAINZ for financial support.
\end{acknowledgements}

\section{Appendix: Tauc gap}

The optical Tauc gap $ \Delta E_{Tauc} $ is a convenient definition for 
the gap of amorphous phases, which circumvents the difficulty that the 
band structure of disordered systems is not properly defined. It relies on the following 
relation \cite{Tauc} between the experimental optical gap $ \Delta E_{gap} $ 
and the optical absorption coefficient $ \alpha $ as a function of 
the photon energy $ h \nu $:
\begin{eqnarray}
\label{alphahnu}
\alpha(h \nu) h \nu \propto (h \nu - \Delta E_{gap})^2, 
\end{eqnarray}
which is applicable to (amorphous) semi-conductors within a certain range of 
photon energies just beyond the gap $\Delta E_{gap}$.
For the on-the-fly calculation of $ \Delta E_{Tauc} $ within each optimization step,
we first compute the optical absorption coefficient $ \alpha $ from 
\cite{Bouzerar}:
\begin{eqnarray}
\label{alpha}
\alpha(h \nu) = \frac{K}{h \nu} \int_{E_F}^{E_F+ h \nu}
n (E-h \nu ) \tilde{n} (E) dE, 
\end{eqnarray}
where $K$ is a constant, while $n$ and $ \tilde{n}$ are the densities of the 
occupied and unoccupied states that are computed
from the eigenvalue spectrum of the DFT Hamiltonian after 
self-consistency has been achieved. Plotting $\sqrt{\alpha(h \nu)~ h \nu}$ as a function of $h \nu$
within a photon energy range around the gap obeys a linear regime, from which $\Delta E_{Tauc} = \Delta E_{gap}$ 
can be obtained by taking the intersection between the linear fit with the horizontal $\sqrt{\alpha(h \nu)~ h \nu} = 0$ axis. 
We note that the value of the constant $K$ is irrelevant 
for the value of $ \Delta E_{Tauc} $ resulting from this approach.
Since the linear behavior only applies to a finite energy range
just beyond the gap, the linear fit has to be restricted to this 
energy interval. 
For the automatic computation of $\Delta E_{Tauc}$, we have selected this interval 
to be within the interval $( (1-\Delta) f W_{tot} , f W_{tot} )$, where $W_{tot}$ is the total width of the spectrum. 


\begin{thebibliography}{90}
\bibitem{Zallen} R. Zallen, {\it The Physics of Amorphous Solids}, Wiley, New York, 1983.
\bibitem{Elliott} S. R. Elliott, {\it Physics of Amorphous Materials}, Longman Scientific \& Technical, Essex, 1990.
\bibitem{Zachariasen} W. H. Zachariasen, J. Am. Chem. Soc. {\bf 54}, 3841 (1932).
\bibitem{Jansen} M. Jansen, J. C. Sch\"on and L. van W\"ullen, Angew. Chem. Int. Ed. {\bf 45}, 4244 (2006).
\bibitem{Maddox} J. Maddox, Nature {\bf 335}, 6187 (1988).
\bibitem{Deaven} D. M. Deaven and K. M. Ho, Phys. Rev. Lett. {\bf 75}, 288 (1995).
\bibitem{WalesScheraga} D. J. Wales and H. A. Scheraga, Science {\bf 285}, 1368 (1999).
\bibitem{Martonak} R. Martonak, A. Laio and M. Parrinello, Phys. Rev. Lett. {\bf 90}, 075503 (2003).
\bibitem{MinHopping} S. Goedecker, J. Chem. Phys. {\bf 120}, 9911 (2004).
\bibitem{USPEX} C. W. Glass, A. G. Oganov and N. Hansen, Comp. Phys. Commun. {\bf 175}, 713 (2006).
\bibitem{Trimarchi} G. Trimarchi and A. Zunger, Phys. Rev. B {\bf 75}, 104113 (2007).
\bibitem{Catlow} S. C. Woodley and R. Catlow, Nature Mater. {\bf 7}, 937 (2008).
\bibitem{Wehmeyer} C. Wehmeyer, G. F. von Rudorff, S. Wolf, G. Kabbe, D. Sch\"arf, T. D. K\"uhne and D. Sebastiani, J. Chem. Phys. {\bf 137}, 194110 (2012).
\bibitem{SA} S. Kirkpatrick, C. D. Gelatt, M. P. Vecchi, Science {\bf 220}, 671 (1983).
\bibitem{CarParrinello} R. Car and M. Parrinello, Phys. Rev. Lett. {\bf 55}, 2471 (1985).
\bibitem{KuhneParrinello} T. D. K\"uhne, M. Krack, F. R. Mohamed and M. Parrinello, Phys. Rev. Lett. {\bf 98}, 066401 (2007).
\bibitem{Pasquarello} J. Sarnthein, A. Pasquarello and R. Car, Phys. Rev. Lett. {\bf 74}, 4682 (2005); Phys. Rev. B {\bf 52}, 12690 (1995).
\bibitem{Marks} N. A. Marks, D. R. McKenzie, B. A. Pailthorpe, M. Bernasconi and M. Parrinello, Phys. Rev. Lett. {\bf 76}, 768 (1996); Phys. Rev. B {\bf 54}, 
9703 (1996).
\bibitem{CaravatiAPL} S. Caravati, M. Bernasconi, T. D. K\"uhne, M. Krack and M. Parrinello, Appl. Phys. Lett. {\bf 91}, 171906 (2007).
\bibitem{CaravatiPRL} S. Caravati, M. Bernasconi, T. D. K\"uhne, M. Krack and M. Parrinello, Phys. Rev. Lett. {\bf 102}, 205502 (2009).
\bibitem{CaravatiJPCM1} S. Caravati, M. Bernasconi, T.D. K\"uhne, M. Krack and M. Parrinello, J. Phys.: Condens. Matter {\bf 21}, 255501 (2009). 
\bibitem{CaravatiJPCM2}  S. Caravati, D. Colleoni, R. Mazzarello, T. D. K\"uhne, M. Krack, M. Bernasconi and M. Parrinello, J. Phys.: Condens. Matter 23, 265801 (2011).
\bibitem{Camellone} M. F. Camellone, T. D. K\"uhne and D. Passerone, Phys. Rev. B {\bf 80}, 033203 (2009).
\bibitem{McGreevy1} R. L. McGreevy and L. Pusztai, Mol. Simul. {\bf 1}, 359 (1988). 
\bibitem{McGreevy2} R. L. McGreevy, J. Phys.: Condens. Matter {\bf 13}, R877 (2001). 
\bibitem{Metropolis} N. Metropolis, A. W. Rosenbluth, M. N. Rosenbluth, A. H. Teller and E. Teller, J. Chem. Phys. {\bf 21}, 1087 (1953).
\bibitem{Kugler} S. Kugler, L. Pusztai, L. Rosta, P. Chieux and R. Bellissent, Phys. Rev B {\bf 48}, 7685 (1993).
\bibitem{Franceschetti} A. Franceschetti and A. Zunger, Nature {\bf 401}, 60 (1999).
\bibitem{Keating} P. N. Keating, Phys. Rev. B {\bf 149}, 674 (1966).
\bibitem{DFT} R. O. Jones and O. Gunnarsson, Rev. Mod. Phys. {\bf 61}, 689 (1989).
\bibitem{Duane} S. Duane, A. D. Kennedy, B. J. Pendleton and D. Roweth, Phys. Lett. B {\bf 2}, 216 (1987).
\bibitem{Ray} J. R. Ray, Phys. Rev. A {\bf 44}, 4061 (1991).  
\bibitem{Los} J. H. Los, L. M. Ghiringhelli, E. J. Meijer and A. Fasolino, Phys. Rev. B {\bf 72}, 214102 (2005).
\bibitem{Ghiringhelli} L. M. Ghiringhelli, J. H. Los, A. Fasolino and E. J. Meijer, Phys. Rev. B {\bf 72}, 214103 (2005).
\bibitem{Tauc} J. Tauc, A. Menth and D. L. Wood, Phys. Rev. Lett. {\bf 25}, 749 (1970).  
\bibitem{Gilkes} W. R. Gilkes, P. H. Gaskell, and J. Robertson, Phys. Rev. B {\bf 51}, 12303 (1995).
\bibitem{VandeVondele} J. VandeVondele, M. Krack, F. Mohamed, M. Parrinello, T. Chassaing and J. Hutter, Comput. Phys. Commun. {\bf 167}, 103 (2005).
\bibitem{PBE} J. P. Perdew, K. Burke and M. Ernzerhof, Phys. Rev. Lett. {\bf 77}, 3865 (1996).
\bibitem{Goedecker} S. Goedecker, M. Teter and J. Hutter, Phys. Rev. B {\bf 54}, 1703 (1996).
\bibitem{Kuehne2009} T. D. K\"uhne, M. Krack and M. Parrinello, \textit{J. Chem. Theory Comput.} \textbf{5}, 235 (2009).
\bibitem{Kuehne2011} T. D. K\"uhne, T. A. Pascal, E. Kaxiras and Y. Jung, \textit{J. Phys. Chem. Lett.} \textbf{2}, 105 (2011).
\bibitem{PascalWater} T. A. Pascal, D. Sch\"arf, Y. Jung and T. D. K\"uhne, J. Chem. Phys. {\bf 137}, 244507 (2012).
\bibitem{Kuehne2013} T. D. K\"uhne and R. Z. Khaliullin, \textit{Nature Commun.} \textbf{4}, 1450 (2013).
\bibitem{MauriNMR1} F. Mauri, B. G. Pfrommer and S. G. Louie, Phys. Rev. Lett. {\bf 77}, 5300 (1996).
\bibitem{MauriNMR2} F. Mauri, B. G. Pfrommer and S. G. Louie, Phys. Rev. Lett. {\bf 79}, 2340 (1997).
\bibitem{SebastianiParrinelloNMR} D. Sebastiani and M. Parrinello, 
J. Phys. Chem. A {\bf 105}, 1951 (2001).
\bibitem{SebastianiParrinelloWaterNMR} D. Sebastiani and M. Parrinello, Phys. Chem. Chem. Phys. {\bf 3}, 675 (2002).
\bibitem{Bouzerar} R. Bouzerar, C. Amory, A. Zeinert, M. Benlahsen, B. Racine,
O. Durand-Drouhin and M. Clin, Journal of Non-Crystalline Solids {\bf 281}, 
171 (2001).
\end{thebibliography}
\end{document}